\documentclass[12pt]{article}
\usepackage{graphicx}

\def\birmin{School of Physics and Astronomy, \\ University of Birmingham, \\ Birmingham, B15 2TT, \\United Kingdom}
\def\support{\footnote{On behalf of the NA62 Collboration}}

\def\Title#1{\begin{center} {\Large #1 } \end{center}}
\def\Author#1{\begin{center}{ \sc #1} \end{center}}
\def\Address#1{\begin{center}{ \it #1} \end{center}}

\newenvironment{Abstract}{\begin{quotation}  }{\end{quotation}}

\def\beq{\begin{equation}}
\def\eeq#1{\label{#1}\end{equation}}
\def\eeqn{\end{equation}}

\def\beqa{\begin{eqnarray}}
\def\eeqa#1{\label{#1}\end{eqnarray}}
\def\eeqan{\end{eqnarray}}

\let\bar=\overbar

\def\Dslash{\not{\hbox{\kern-4pt $D$}}}
\def\dslash{\not{\hbox{\kern-2pt $\del$}}}

\def\msb{{\bar{\ssstyle M \kern -1pt S}}}

\begin{document}
\begin{titlepage}

\vfill
\Title{Lepton universality tests with leptonic kaon decays}
\vfill
\Author{ Cristina Lazzeroni\support}
\Address{\birmin}
\vfill
\begin{Abstract}
The NA62 experiment at CERN aims to use rare kaon decays to search for new phenomena beyond the Standard Model.
During the current short term phase (data taking completed in years 2007-8), the ratio $R_K = \Gamma(K^+ \rightarrow e \nu_e (\gamma)) / 
\Gamma(K^+ \rightarrow \mu \nu_{\mu} (\gamma))$ of leptonic decay rates is studied, which tests the structure of weak interactions and lepton flavour universality. In this paper, the $R_K$ analysis is summarized, and the preliminary result is discussed, based on 59963 $K^+ \rightarrow e \nu_e $ candidates collected in 2007. 
\end{Abstract}
\vfill
\begin{center}
Proceedings of CKM2010\\
the 6th International Workshop on the CKM Unitarity Triangle\\
University of Warwick, UK, 6-10 September 2010
\end{center}
\vfill
\end{titlepage}
\def\thefootnote{\fnsymbol{footnote}}
\setcounter{footnote}{0}
%




The ratio of kaon leptonic decay rates $R_K = \Gamma(K^+ \rightarrow e^+ \nu_e) / 
\Gamma(K^+ \rightarrow \mu^+ \nu_{\mu})$ has been calculated with an excellent accuracy within the SM\cite{ciri1}: $R_K(SM)=(2.477 \pm 0.001) \times 10^{-5}$.


The ratio $R_K$ is sensitive to lepton flavour universality violation (LFV) effects originating at
one-loop level from $H^\pm$ exchange in two-Higgs-doublet models \cite{mas1,mas2}, and the mixing effects in the right-handed slepton sector, providing a unique probe into this aspect of supersymmetric flavour physics\cite{ellis}. $R_K$ receives the following leading-order contribution due to LFV coupling of the Higgs boson:

$$\frac{ \Delta R_K} {R^{SM}_K} = (M_K / M_H)^4 (M_{\tau} /M_e)^2 |\Delta^{31}_R|^2 tan^6 \beta  $$

where $\Delta^{31}_R \sim 10^{-3}$ is the mixing parameter between the superpartners of the right-handed leptons. This can enhance $R_K$ by O(1\%) for large $tan \beta$ and $M_H$ 
(for example $tan \beta=40$ and $M_H$=500 GeV)\cite{mas1}.
The current world average (including only final results, and thus ignoring the preliminary
NA48/2 ones) is $R^{WA}_K = (2.490 \pm 0.030) \times 10^{-5}$, dominated by a recent measurement by the KLOE collaboration\cite{ambro}. 

\section{Beam, detector and data taking}
The NA62 experiment at CERN collected a dedicated data sample in
2007-08, aiming at a measurement of $R_K$ with a 0.4\% precision.
The present analysis is based on 40\% of the data sample.
The beam line and setup of the NA48/2 experiment\cite{na48} were used for the NA62  data taking. Experimental conditions and trigger logic were optimized for the $R_K$ measurement.
The beam line delivered simultaneous unseparated $K^+$ and $K^-$ beams derived
from 400 $GeV/c$ primary protons extracted from the CERN SPS. Most of the data, including
the sample used for the present analysis, were collected with the $K^+$ beam only, as the muon
sweeping system provided better suppression of the positive beam halo component. A narrow
momentum band of $(74.0\pm1.6) GeV/c$ was used to minimize the corresponding contribution to
resolution in kinematical variables.
The fiducial decay region was contained in a 114 $\rm{m}$ long cylindrical vacuum tank. The beam
flux at the entrance to the decay volume was $2.5 \times 10^7$ particles per pulse of 4.8 $\rm{s}$ duration. 
The fractions of $K^+ , \pi^+ , p, e^+ , \mu^+$ in the beam were 0.05, 0.63, 0.21, 0.10, 0.01 respectively. The fraction of beam kaons decaying in the vacuum tank at nominal momentum was 18\%. 
The transverse size of the beam within the decay volume was $ x =  y = 7$ $\rm{mm}$ (rms), and its angular divergence was negligible.
A minimum bias trigger configuration was employed, resulting in high efficiency with relatively
low purity. The $Ke2$ trigger condition consisted of coincidence of hits in the plastic scintillator hodoscope
with $10 GeV$ 
energy deposition in the calorimeter. The $K\mu2$ trigger condition consisted of the 
hodoscope signal alone downscaled by a factor of 150. 

\section{Analysis strategy and event selection}
The analysis strategy is based on counting the numbers of reconstructed $Ke2$ and $K\mu2$ candidates collected at the same time. Consequently the result does not rely on kaon 
flux measurement, and several systematic effects cancel to first order.
To take into account the significant dependence of signal acceptance and background level
on lepton momentum, the measurement is performed independently in bins of this observable:
10 bins covering a lepton momentum range of $[13, 65] GeV/c$ are used, where the first bin spans 7 GeV while the others are 5 GeV wide. 



A detailed Monte Carlo (MC) simulation including beam line optics, full detector geometry
and material description, stray magnetic fields, local inefficiencies of chamber wires, and time variations of the above throughout the running period, was used to evaluate the acceptance correction
and the geometric parts of the acceptances for background processes. 
Simulations are used to a
limited extent only: particle identification, trigger and readout efficiencies are measured directly using data.

Due to the topological similarity of $Ke2$ and $K\mu2$ decays, a large part of the selection conditions
is common for both decays\cite{evgueni}.
The following two principal selection criteria are different for the $Ke2$ and $K\mu2$ decays: 
(1) the kinematic identification is based on the reconstructed squared missing mass assuming the track
to be a positron or a muon; (2) particle identification is based on the ratio $E/p$
of track energy deposit in the calorimeter to its momentum measured by the spectrometer.
The quantity $M^2_{miss} = (P_K - P_l)^2$ is defined, where $P_K$ and $P_l$ ($l = e, \mu$) are the four-momenta of the kaon (average beam momentum assumed) and the lepton (positron or muon
mass assumed). A selection condition $-M^2_1 < | M^2_{miss} (e, \mu)| < M^2_2$ is applied, where $M^2_1 , M^2_2$ vary in the ranges $0.013-0.016$ and $0.010-0.014$ (GeV/c$^2)^2$ respectively, optimised taking into account resolution and backgrounds.
Particles with $(E/p)_{min} < E/p < 1.1$, where $(E/p)_{min} = 0.95$ for 
$p > 25$ GeV/c and  $(E/p)_{min} = 0.90$ otherwise, are identified as positrons. Tracks with
$E/p < 0.85$ are identified as muons.

\begin{figure}[t]
\begin{center}
\includegraphics[width=0.45\textwidth]{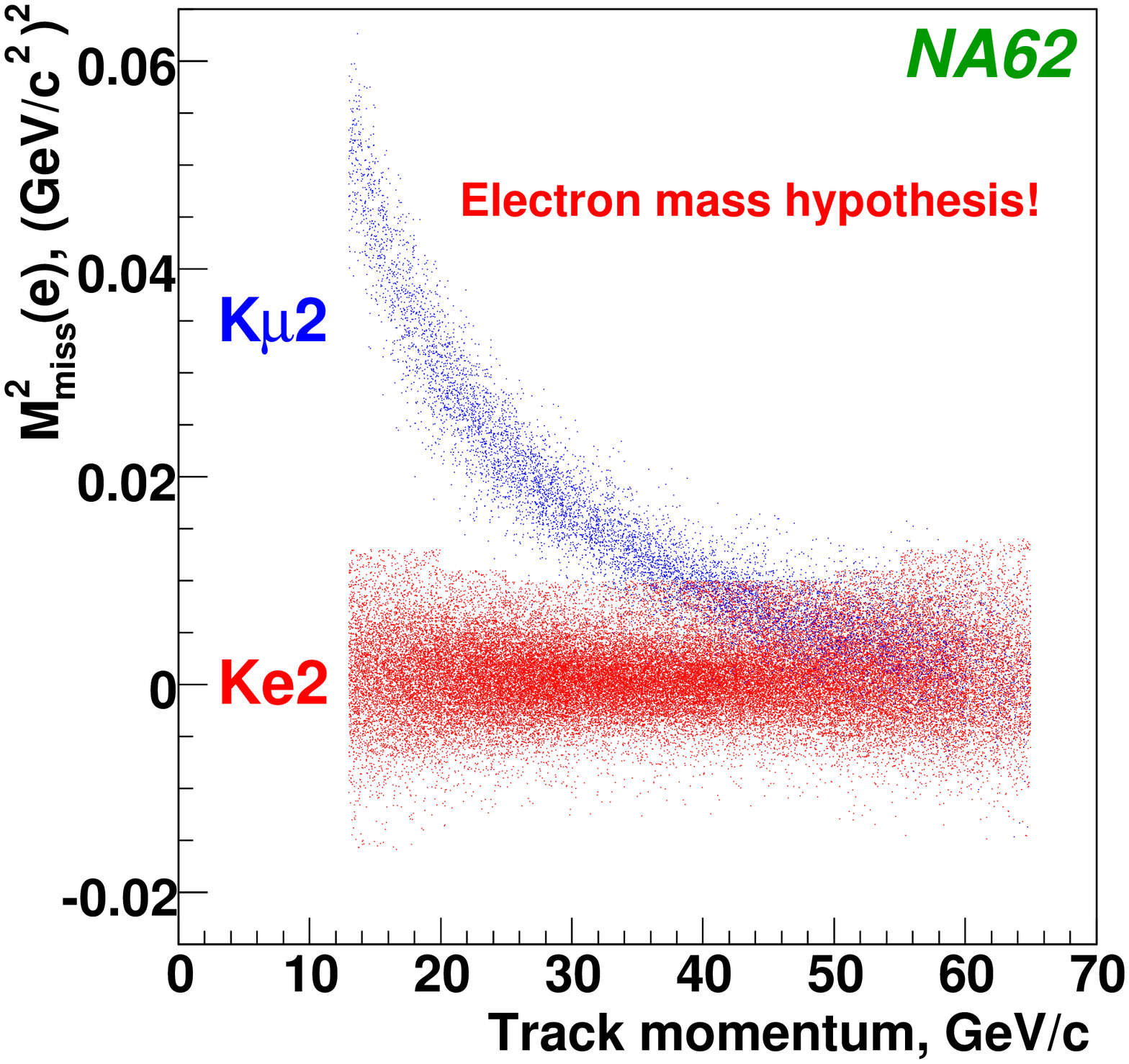}
\hspace{0.02\textwidth}
\includegraphics[width=0.42\textwidth]{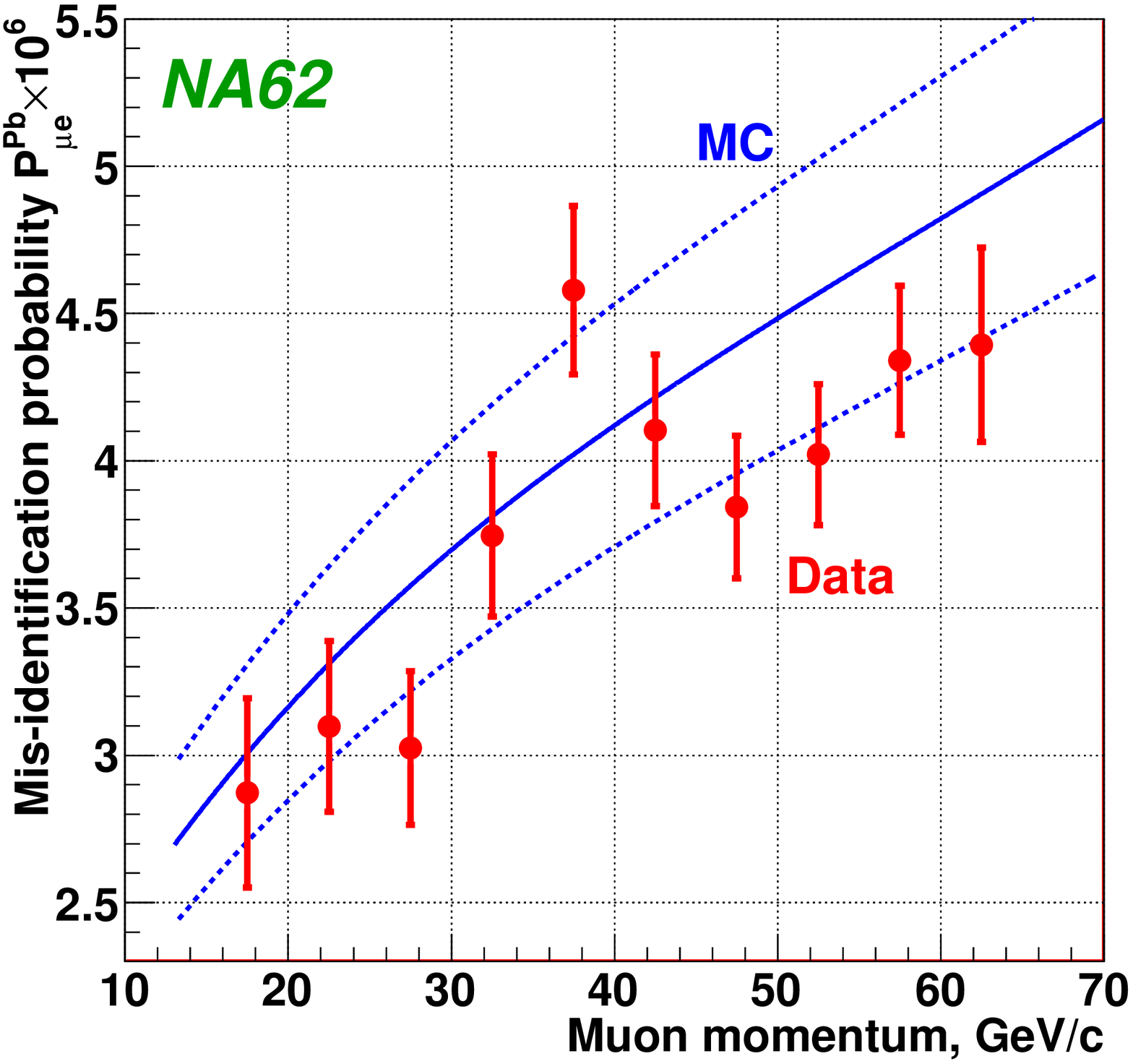}
\caption{(Left) Missing mass squared in positron hypothesis vs lepton momentum for reconstructed $Ke2$ and $K\mu2$ decays: kinematic separation is possible at low lepton momentum only. (Right) Muon mis-ID probability vs track momentum (Pb wall installed): data and MC simulation.}
\label{backg}
\end{center}
\end{figure}

\section{Background rejection}
$K\mu2$ decays with a mis-identified muon are the main background source in the $Ke2$ sample.
Sufficient kinematic separation between electron and muon decays is not achievable at high lepton momentum ($p > 30 GeV/c$), as shown in Fig.\ref{backg}(Left). The probability of muon identification as positron in that momentum range is $P(\mu \rightarrow e) \sim 4 \times 10^{-6}$, as $E/p > 0.95$ due to catastrophic bremsstrahlung in or in front of the calorimeter.
Since such probability is non-negligible for the measurement in question, a direct
measurement of $P(\mu \rightarrow e)$ to $10^{-2}$ relative precision is necessary to validate the theoretical calculation of the bremsstrahlung cross-section\cite{kel} in the high energy $\gamma$ range used to evaluate the $K\mu2$ background.
The available muon samples are typically affected by $\sim 10^{-4}$ electron/positron contamination
due to $\mu \rightarrow e$ decays in flight, which obstructs the probability measurements.
In order to obtain sufficiently pure muon samples, a $9.2 X^0$ thick lead (Pb) wall covering   
20\% of the geometric
acceptance was installed in front of the calorimeter during
a fraction of the data taking. In the samples of tracks traversing the Pb and having $E/p > 0.95$,
the electron component is suppressed to a level of $\sim 10^{-8}$ by energy losses in Pb.

However, Pb wall modifies $P(\mu \rightarrow e)$ via two principal mechanisms: 1) muon energy loss in the Pb by ionization, dominating at low momentum; 2) bremsstrahlung in Pb, dominating at high momentum. To evaluate the corresponding correction factor $f_{Pb} = P_{\mu e} / P^{Pb}_{\mu e} $, a dedicated Geant4 based simulation of muon propagation downstream the spectrometer
involving all electromagnetic processes, including muon bremsstrahlung has been developed.
The relative systematic uncertainties on $P_{\mu e}$ and $P^{Pb}_{\mu e} $ obtained from simulation are estimated to be 10\% - measured and simulated probabilities are shown 
Fig.\ref{backg}(Right). The error affecting their ratio is significantly smaller (2\%) due to partial cancellation of uncertainties. The $K\mu2$ background has been computed to be $(6.10\pm0.22)$\% using the measured $P^{Pb}_{\mu e} $ corrected for the simulated $f_{Pb}$ and correcting for
the correlation between the reconstructed $M^2_{miss} (e)$ and $E/p$.




\begin{figure}[htbp]
\begin{center}
\includegraphics[width=0.45\textwidth]{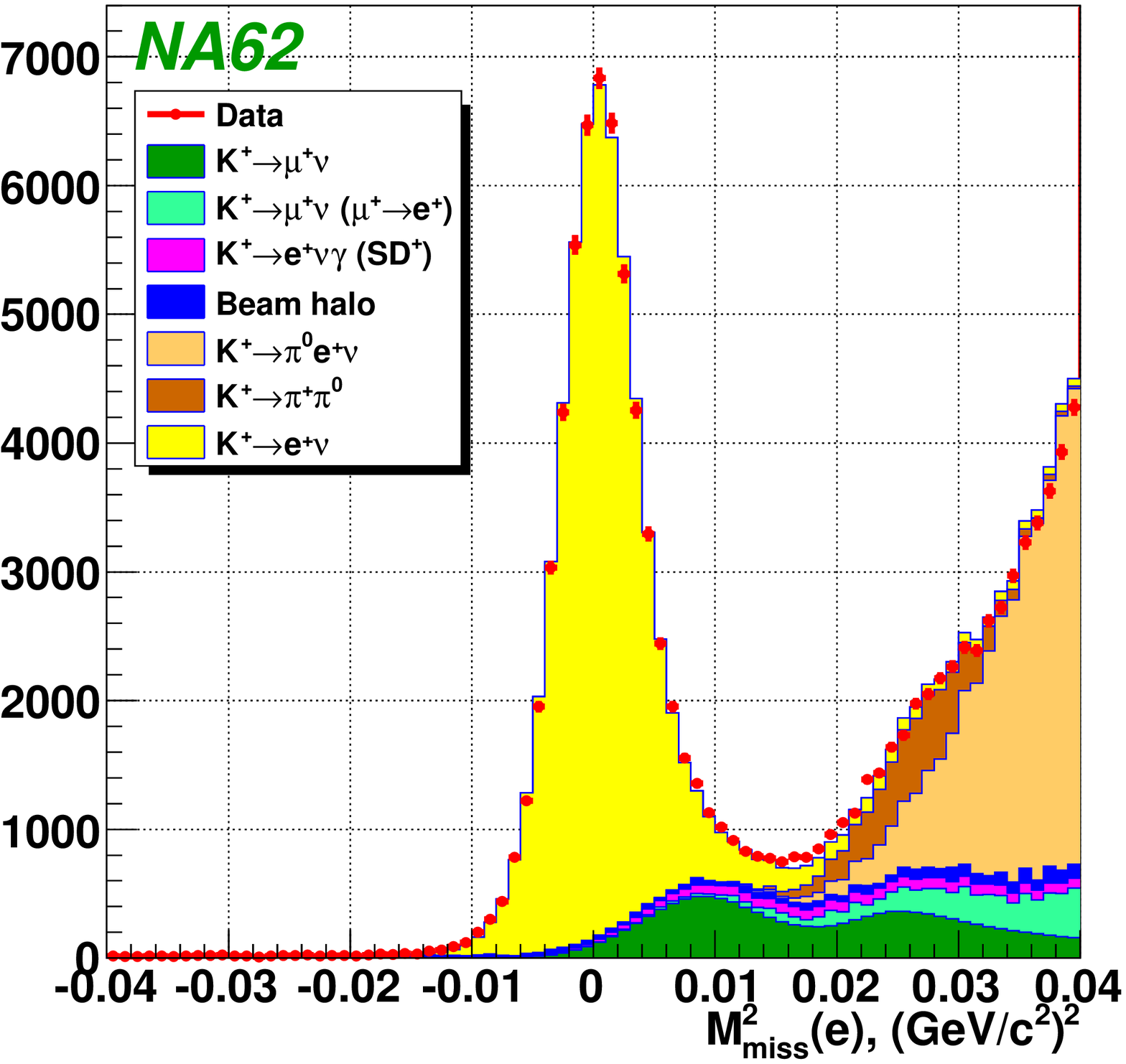}
\hspace{0.02\textwidth}
\includegraphics[width=0.45\textwidth]{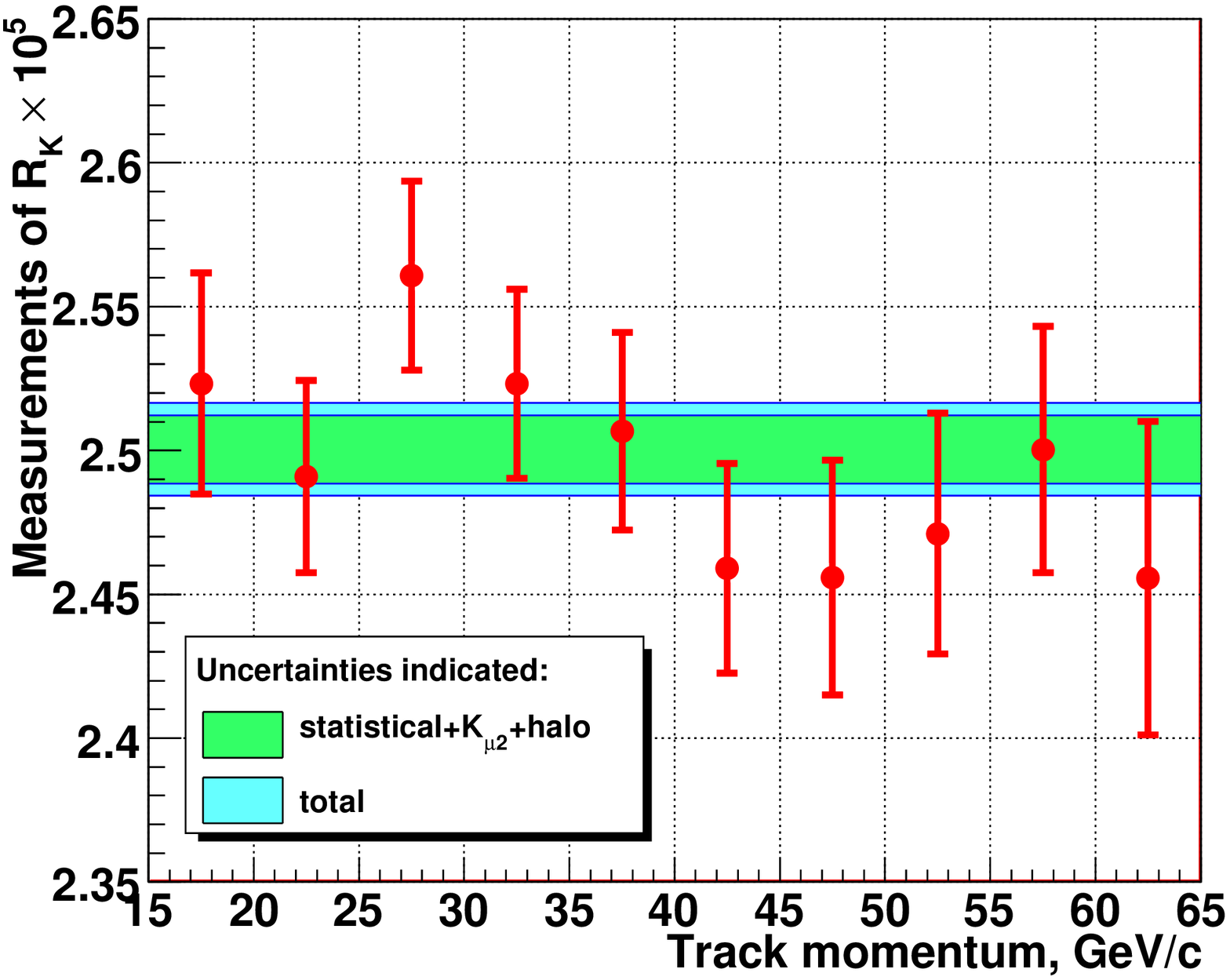}
\caption{(Left) Reconstructed squared missing mass distribution for the $Ke2$ candidates: data (dots)
presented as sum of MC signal and background contributions (filled areas). (Right) Measurements of $R_K$ in lepton momentum bins.}
\label{result}
\end{center}
\end{figure}

The number of candidates is $N(Ke2) = 59963$ 
and $N(K\mu2) = 1.803\times 10^7$. The missing mass distributions (electron hypothesis) of data events and backgrounds are presented in Fig.\ref{result}(Left). 
Backgrounds integrated over lepton momentum are
summarized in Table \ref{bkg}.
The total background is $(8.78\pm0.29)$\%.

\begin{table}[htdp]
\begin{center}
\begin{tabular}{|cc|cc|cc|}
\hline
Source  & $N_B / N_{tot}$(\%) & Source &  $N_B / N_{tot}$(\%) & Source & $N_B / N_{tot}$(\%) \\
\hline
$K\mu2$  &  $6.10\pm0.22$ &  $Ke2\gamma$(SD) & $1.15\pm0.17$ & $Ke3$ &  $0.06\pm0.01$ \\
$K\mu2 (\mu \rightarrow e)$  &  $0.27\pm0.04$ & Beam halo & $1.14\pm0.06$ & $K2\pi$ & $0.06\pm0.01$\\
\hline
\end{tabular}
\caption{Summary of background sources in $Ke2$ sample.}
\label{bkg}
\end{center}
\end{table}

\section{ Systematic uncertainties and result}
Positron identification efficiency is measured directly as a function of momentum and calorimeter
impact point, using pure samples of electrons and positrons obtained by kinematic selection of
$K^+ \rightarrow  \pi^0 e^+ \nu$ decays collected concurrently with the $Ke2$ sample, and $K^0_L
\rightarrow \pi^\pm e^\pm \nu$  decays from a special $K^0_L$ run. The $K^+$ and $K^0_L$ measurements are in good agreement. The measured $f_e$ averaged over the $Ke2$ sample is 
$1- f_e = (0.73 \pm 0.05)$\%.
Muon identification inefficiency is negligible.
The radiative $K^+ \rightarrow e^+ \nu \gamma$ (IB) process is simulated following\cite{ecker} with higher order corrections according to \cite{weinb,gatti}.
An additional systematic uncertainty
reflects the precision of beam line and apparatus description in the MC simulation.
Trigger efficiency correction $1 - \epsilon= (0.41\pm0.05)$\% accounts for the fact that the condition
$E_{calo} > 10 GeV$ is used in the $Ke2$ trigger only. 
A conservative systematic uncertainty 
is assigned due to effects of trigger dead time which affect the two modes differently. Calorimeter global
readout efficiency $f_{calo}$ is measured directly to be $1 - f_{calo} = (0.20 \pm 0.03)$\%
and is checked to be stable in time using an independent readout system.
The independent measurements of $R_K$ in lepton momentum bins, and the result combined
over the momentum bins are presented in Fig.\ref{result}(Right). 
The preliminary result, based on 40\% of data, is $R_K = (2.486\pm0.011_{stat.} \pm0.008_{syst.})\times 10^{-5} = (2.486\pm 0.013) \times 10^{-5}$, and is consistent with Standard Model expectations.



\end{document}